\documentclass[conference]{IEEEtran}
\IEEEoverridecommandlockouts

\usepackage[bookmarks=false]{hyperref}

\usepackage{orcidlink}
\usepackage{cite}
\usepackage{amsmath,amssymb,amsfonts}
\usepackage{algorithmic}
\usepackage{graphicx}
\usepackage{textcomp}
\usepackage{xcolor}
\usepackage{comment}
\usepackage{array}
\usepackage{tabularx}
\usepackage{booktabs} % For formal tables
\usepackage{listings}
\usepackage{mathtools} % For multiline equations
\usepackage[T1]{fontenc}
\usepackage{amsmath,amssymb}
\usepackage{amsfonts}
\usepackage{pifont}
\usepackage{units}
\usepackage{regexpatch}
\makeatletter
\xpatchcmd{\@todo}{\setkeys{todonotes}{#1}}{\setkeys{todonotes}{inline,#1}}{}{}
\makeatother

\def\BibTeX{{\rm B\kern-.05em{\sc i\kern-.025em b}\kern-.08em
    T\kern-.1667em\lower.7ex\hbox{E}\kern-.125emX}}


\begin{document}

\title{Decentralized Credential Status Management: \\ A Paradigm Shift in Digital Trust}

\author{\IEEEauthorblockN{Patrick Herbke}
\IEEEauthorblockA{\textit{Service-centric Networking} \\
\textit{Technische Universität Berlin}\\
Berlin, Germany \\
p.herbke@tu-berlin.de}
\and
\IEEEauthorblockN{Thomas Cory}
\IEEEauthorblockA{\textit{Service-centric Networking} \\
\textit{Technische Universität Berlin}\\
Berlin, Germany \\
cory@tu-berlin.de}
\and
\IEEEauthorblockN{Mauro Migliardi}
\IEEEauthorblockA{\textit{Department of Electronic Engineering} \\
\textit{University of Padua}\\
Padua, Itlay \\
mauro.migliardi@unipd.it}}

\maketitle

\begin{abstract}
Public key infrastructures are essential for Internet security, ensuring robust certificate management and revocation mechanisms. The transition from centralized to decentralized systems presents challenges such as trust distribution and privacy-preserving credential management. The shift to decentralized systems addresses single points of failure in centralized systems and leverages the transparency and resilience of decentralized technologies. This paper explores the evolution of certificate status management from centralized to decentralized frameworks, focusing on blockchain technology and advanced cryptography. We provide a taxonomy of the challenges of centralized systems and discuss opportunities provided by existing decentralized technologies. Our findings reveal that, although blockchain technologies enhance security and trust distribution, they represent a bottleneck for parallel computation and face inefficiencies in cryptographic computations. For this reason, we propose a framework of decentralized technology components that addresses such shortcomings to advance the paradigm shift toward decentralized credential status management. 
\end{abstract}


\section{Introduction} 
\label{sec:introduction} 
Public key infrastructure (PKI) faces challenges such as single points of failure and trust monopolies, as evidenced by breaches like DigiNotar~\cite{prins2011diginotar}, Symantec~\cite{DBLP:conf/pam/GasserHHKHC18}, and more recently Equifax~\cite{daswani2021equifax}. Decentralized Credential Status Management (DCSM) manages credential status by shifting digital trust from centralized PKI (CPKI) to decentralized PKI (DPKI) using blockchain, multiparty computation (MPC), fully homomorphic encryption (FHE), and zero-knowledge proofs (ZKP).

Trust, digital trust, and decentralized trust are foundational concepts in the security and reliability of digital systems, particularly within decentralized environments~\cite{9663537}. Trust refers to the confidence in the integrity, reliability, and security of systems and entities. Digital trust encompasses this confidence within digital interactions, relying on cryptographic technologies and protocols. Decentralized trust distributes the trust anchor among multiple nodes or entities, enhancing resilience and mitigating single points of failure.

This paper explores the transition from centralized certificate management to decentralized credential management, highlighting their differences and implications for security and efficiency. Traditionally managed within CPKI, certificates are digital attestations that bind a public key to an entity's identity, typically issued and verified by certificate authorities (CAs). In contrast, Verifiable Credentials (VCs) in a DPKI encompass a broader range of verifiable claims about an entity, not limited to public key bindings, but also including attributes like qualifications, roles, or statuses. Shifting from CPKI to DPKI enhances security by leveraging distributed ledger technology's inherent properties of immutability and transparency. A DPKI distributes trust across multiple nodes, reducing single points of failure and increasing resilience to attacks by making it harder for malicious actors to compromise the entire system. Furthermore, the decentralized nature of DPKI ensures that no single entity controls the entire infrastructure, further reducing vulnerability to breaches and improving overall trust in the system~\cite{DBLP:conf/sp/MatsumotoR17}. 

Our research contributes to the management of trust and credential status in decentralized networks. We provide insights into managing digital trust and credentials by introducing a taxonomy that outlines the limitations of CPKI and explore opportunities within DPKI systems. The research questions addressed are: (1) How can decentralized technologies enhance credential status management? (2) What cryptographic primitives and governance frameworks are required for robust DCSM? (3) How can DCSM be effectively integrated into the Self-Sovereign Identity (SSI)~\cite{preukschat2021self} paradigm to improve user-centric and privacy-preserving identity management?

This paper offers a detailed comparative analysis of traditional PKI components and decentralized approaches. We summarize the findings as a technology framework that serves as a foundation for future research and innovation in this critical area.

\section{Evolution of Public Key Infrastrucure}
PKI secures digital communications by managing digital certificates that authenticate public keys. This section traces the evolution of PKI from centralized to decentralized systems.

\subsection{Foundational Cryptographic Primitives and Standards} 
Key cryptographic advancements in the 1970s and 1980s, such as the Diffie-Hellman protocol~\cite{1055638}, RSA~\cite{10.1145/359340.359342}, and Merkle's puzzle~\cite{merkle22method}, laid the foundation for secure communication. Goldwasser et al.~\cite{DBLP:conf/stoc/GoldwasserMR85} introduced zero-knowledge proofs, essential for privacy-preserving cryptographic protocols.
The X.509 standard~\cite{x509} marked a milestone for PKI, defining digital certificate formats and the role of CAs. Furthermore, Chaum's work~\cite{chaum1983blind} on privacy-preserving transactions and decentralized digital cash introduced the proto-blockchain protocol.

However, adopting these decentralized systems faced societal, technological, and integration challenges. Developments in the 1970s and 1980s laid the foundation for cryptography and PKI, paving the way for DPKI systems.

\subsection{Security Protocols and the Rise of Decentralized Identity Management}
The 1990s established SSL/TLS for secure data exchange~\cite{dierks1999tls}. Scalability and latency challenges with Certificate Revocation Lists (CRL) led to the Online Certificate Status Protocol (OCSP)~\cite{DBLP:conf/fc/McDanielR00} in 1991. The OCSP introduced network latency and privacy concerns, highlighting limitations in scalability, privacy, and resilience~\cite{DBLP:conf/mm/Wohlmacher00}.

In 1991, Pretty Good Privacy (PGP) emerged as a decentralized alternative, focusing on user-controlled cryptographic keys and direct trust management through the Web of Trust~\cite{DBLP:books/daglib/0084694}. The PGP trust model laid the foundation for future decentralized systems, improving digital trust and identity management. From the 1970s to the 2000s, PKI evolved by enhancing centralized and pioneering decentralized systems that prioritized user control and privacy. This period set the stage for exploring DPKI systems that address the challenges of centralized models.

\subsection{Advancements in Centralized and Decentralized PKI Systems}
The 2000s introduced innovations in centralized PKI and certificate management. The Automatic Certificate Management Environment simplified the management of digital certificates, but presented challenges in wide adoption and security~\cite{Barnes2019AutomaticCM}. Certificate Transparency introduced public certificate logs to increase trust and mitigate risks such as CA compromises and man-in-the-middle attacks~\cite{10.1145/2659897}. However, the challenges in certificate verification and revocation information remained~\cite{10.1145/2815675.2815685}.

In the 2000s, decentralized technologies advanced with Hal Finney's Reusable Proof of Work~\cite{finney2004rpow} and the introduction of Bitcoin~\cite{nakamoto2008bitcoin} in 2009, demonstrating the potential of blockchain for secure, decentralized systems.

\subsection{Blockchain Era} 
From 2010 to 2024, the blockchain era advanced DPKI and digital identity management. Smart contracts~\cite{DBLP:journals/fgcs/ZhengXDCCWI20}, Decentralized Autonomous Organizations (DAO)~\cite{DBLP:journals/tcss/WangDLYOW19} and Self-Sovereign Identity (SSI)~\cite{tobin2016inevitable} revolutionized the management and governance of digital credentials. SSI enable individuals to control their digital identities through Decentralized Identifiers (DID)~\cite{reed2020decentralized} and VCs~\cite{DBLP:journals/corr/abs-2006-04754}.

The blockchain era marks a shift towards decentralizing digital identity and PKI systems, emphasizing user control, privacy, and advanced cryptographic technologies. However, to realize the full potential of DPKI, it is crucial to address the persistent challenges of CPKI systems and the computational challenges inherent in DPKI.

\section{Related Work}
This section provides a comprehensive overview of existing literature on decentralized credential management systems, focusing on their evolution, challenges, and technological advancements. The related work section examines DPKI models, highlighting their strengths, limitations, and privacy considerations. Additionally, this section reviews existing surveys on trust-based systems. This structured approach establishes a solid foundation for this paper's subsequent discussion and analysis. Furthermire, this section reviews literature on decentralized PKI and Self-Sovereign Identity. 

In 2014, Fromknecht et al.~\cite{Fromknecht2014ADP} proposed Certcoin, a blockchain-based DPKI without central failure points inherent in traditional certificate authority models, leveraging the decentralized nature of blockchain technology. Certcoin focuses on linking identity with public keys in a public ledger, ensuring transparency and security. However, it does not address privacy concerns, as identities and public keys are publicly linked, making it unsuitable for use cases requiring privacy. Unlike our proposed technology framework, Certcoin does not incorporate the SSI, MPC, or FHE principles for privacy-preserving credential transactions. 

Our proposed framework leverages decentralized technologies and advanced cryptographic techniques to enhance security, privacy, and efficiency in credential management. Unlike Certcoin, which links identities and public keys on a public ledger, our framework uses ZKP to allow users to prove possession of a credential without revealing their identity. This approach ensures that user identity remains private while providing verifiable proof of credential ownership. Additionally, MPC allows multiple parties to compute functions over their inputs without revealing the inputs themselves, enhancing privacy during credential issuance and verification. FHE enables computations on encrypted data, ensuring that sensitive information remains confidential. These techniques collectively address privacy concerns associated with publicly linking identities and public keys, providing a secure and privacy-preserving solution.

Axon and Goldsmith~\cite{axon2016pb} presented PB-PKI in 2016, which adapts Certcoin to privacy-sensitive areas by unlinking identities from public keys on the blockchain. PB-PKI introduces user-controlled disclosure and unlinkable short-term key updates to address privacy concerns in IoT applications. Although PB-PKI improves privacy, it does not integrate SSI concepts for user-centric identity management. Furthermore, PB-PKI lacks advanced cryptographic techniques, such as MPC and FHE, which are essential for secure and privacy-preserving computations in decentralized systems. 

Compared to PB-PKI, which addresses privacy by unlinking identities from public keys, our proposed framework further enhances privacy by integrating MPC and FHE. MPC enables multiple entities to collaboratively process credential information without exposing their private data, ensuring that sensitive details remain confidential. FHE allows computations on encrypted data, providing an additional layer of security and privacy. By combining MPC and FHE, our framework ensures that credential transactions are secure and private, mitigating risks associated with data exposure. A combined integration of MPC and FHE represents an advancement over PB-PKI, offering robust privacy protection while maintaining the integrity and verifiability of credentials.

The study by Vanin et al.~\cite{vanin2022blockchain} introduces an FHE approach to ensure end-to-end data protection in healthcare. The proposed system of Vanin et al. aims to mitigate privacy risks and enhance data interoperability among health institutions by utilizing distributed hash tables (DHT) and the InterPlanetary File System (IPFS) for data storage and sharing. The model's effectiveness is evaluated by its ability to protect personal privacy while enabling encrypted data calculations without exposing sensitive information. Our work is distinct from the study by Vanin et al. as we concentrate explicitly on decentralized credential status management.

Similarly, CertLedger is a blockchain-based system for transparent and secure certificate management and revocation proposed by Kubilay et al.~\cite{kubilay2019certledger}. CertLedger focuses on removing single points of failure associated with CAs and ensuring the integrity of the certificate lifecycle through a decentralized ledger. However, like Certcoin, PB-PKI, and the research of Vanin et al., CertLedger does not incorporate SSI principles or advanced cryptographic techniques such as MPC and FHE, limiting its ability to provide user-centric and privacy-preserving identity management solutions.

Fragkos et al.~\cite{9201400} explore artificial intelligence approaches to enhance the performance and privacy of electronic cash systems. Fragkos et al. introduce PUF-Cash, an e-Cash scheme leveraging Physical Unclonable Functions (PUFs) for authentication and encryption between users, banks, and multiple trusted third parties. The AI methods utilized in PUF-Cash optimize the distribution of withdrawal amounts among trusted third parties to maximize privacy and performance. The scheme provided by Fragkos et al. aims to provide strong privacy guarantees, ensuring that users can conduct transactions anonymously, thereby preserving the anonymity implicit in traditional fiat money.

\subsection*{Comparison of CertLedger and PUF-Cash}

CertLedger focuses on linking identity with public keys in a public ledger, ensuring transparency and security. However, it does not address privacy concerns, as identities and public keys are publicly linked, making it unsuitable for use cases requiring privacy. Unlike CertLedger, which emphasizes blockchain transparency over user-centric privacy mechanisms, Fragkos et al. leverages AI and PUFs to enhance privacy and performance in e-Cash transactions. PUF-Cash uses multiple trusted third parties and advanced cryptographic techniques to ensure the anonymity and security of electronic transactions.

Our proposed framework integrates SSI concepts to enhance user-centric and privacy-preserving identity management compared to CertLedger and PUF-Cash. SSI principles, including DIDs and VCs, grant users control over their digital identities without reliance on centralized authorities. Unlike CertLedger, which relies on blockchain transparency, and PUF-Cash, which employs AI for privacy, our approach leverages SSI principles to prevent unauthorized access and profiling, offering a more robust and privacy-focused solution for identity management.

Conversely, Mühle et al.~\cite{muhle2018survey} provide a comprehensive survey outlining SSI's principles, benefits, and challenges. SSI emphasizes user control over digital identities through DIDs and VC, enabling users to manage their credentials without relying on centralized authorities. The survey emphasizes privacy and user autonomy but lacks technical details on integrating SSI with DPKI. Furthermore, the survey does not address using advanced cryptographic techniques such as MPC and FHE.

\section{Transition from Centralized PKI to Decentralized PKI}
Transitioning from CPKI to DPKI marks a fundamental shift in digital security. Centralized systems face single points of failure and complex certificate management, while decentralized systems distribute trust and improve resilience against attacks. This section explores the operational and organizational challenges of the CPKI and DPKI frameworks and outlines conditions for successful implementation.

\begin{figure*}[!ht]
    \centering
    \includegraphics[width=0.9\linewidth]{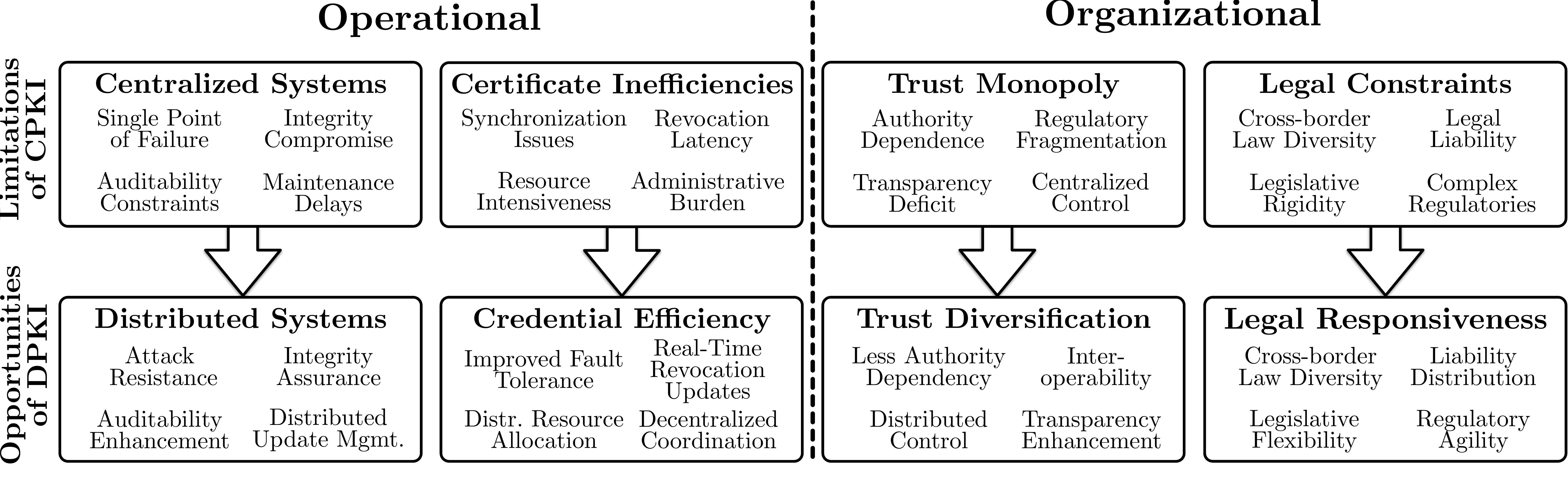}
    \caption{Taxonomy of Operational and Organizational challenges in CPKI and the transition to improvements with DPKI. The figure categorizes the limitations of CPKI, including operational inefficiencies and trust monopolies, and highlights DPKIs' enhancements, such as scalability, improved update efficiency, and diversified trust management.}
    \label{fig:taxonomy-central}
\end{figure*}

\subsection{Operational Considerations}

Operational considerations highlight the limitations of CPKI and the potential changes of DPKI. As depicted on the left side of Figure~\ref{fig:taxonomy-central}, these considerations encompass systemic vulnerabilities and efficiency challenges.

\subsubsection*{Centralization and Certificate Inefficiencies}

CPKI systems face operational limitations due to the centralization of certificate issuance by a few CAs. This centralization introduces single points of failure, making the system vulnerable to attacks~\cite{217541}. The DigiNotar breach illustrated how the compromise of a single CA could result in the widespread issuance of fraudulent certificates, severely affecting multiple high-profile domains. Centralized PKI systems face maintenance delays that require manual intervention for updates, increasing the risk of exploitation. Decentralized solutions can mitigate some risks by distributing control, improving resilience, and reducing single points of failure. However, they do not guarantee a timely patch application, especially for OS-level vulnerabilities or issues requiring a hard fork, which challenge quick resolution. These maintenance delays are further complicated by the administrative burden involved in the certificate lifecycle, including issuance, renewal, and revocation. Revocation latency in OCSP exposes users to potential attacks using revoked certificates~\cite{chung2018web}. 

\subsubsection*{Distribution and Credential Efficiency}

DPKI systems promise to address the operational limitations of CPKI by decentralizing operations, thus improving attack resistance and ensuring data immutability~\cite{qin2020cecoin}. By distributing credential operations such as issuance, revocation, and verification across multiple nodes, DPKI systems mitigate single points of failure and reduce the risk of system-wide compromise. Blockchain technology records all transactions and updates in an immutable ledger, making data alteration or manipulation by malicious actors nearly impossible. However, immutability does not address vulnerabilities within the decentralized system itself. If a malicious actor exploits a vulnerability in the ledger to generate a false certificate, the immutability of the blockchain would not prevent the initial compromise, but would only ensure that the event record remains unchanged~\cite{zheng2017overview}. 

However, DPKI systems face scalability and performance challenges. Although DPKI offers potential advantages such as load distribution across multiple nodes, they have not yet achieved horizontal scaling~\cite{papageorgiou2020dpki}. Current DPKI implementations struggle to maintain an increasing volume of transactions without performance degradation, indicating prevalent scalability bottlenecks. The enhancement of revocation through real-time updates remains an open issue, as demonstrated by verifiable self-sovereign identities offline~\cite{9343191}. Transaction latency issues further reduce operational efficiency. Blockchain-based systems are limited by their inherently sequential processing architecture, unlike CPKI systems, which use parallel computing to improve performance~\cite{xu2021slimchain}. Future research must address these critical scalability and efficiency challenges to realize the full potential of DPKI systems.

\subsection{Organizational Considerations}

Organizational considerations emphasize trust and legal responsibilities~\cite{pop00001}. The shift from CPKI to DPKI promises improvements in compliance management and the adoption of evolving legal requirements. Furthermore, decentralization of legal liabilities enables a collaborative and consortium-driven environment that improves security and equity, cultivating a robust and trustworthy cross-border digital infrastructure~\cite{geneiatakis2020blockchain}.

\subsubsection*{Trust Monopoly in Centralized Systems}

CPKI systems concentrate trust in a few CAs, creating systemic risks. Reliance on limited authorities makes the system vulnerable to single points of failure, where one CA's compromise can undermine the entire network~\cite{braun2013potential}.

\subsubsection*{Trust Diversification with Decentralized Systems}

DPKI reduces reliance on any single authority by distributing trust across multiple entities, enhancing the system's resilience to attacks and failures. Blockchain technology supports trust decentralization by maintaining a transparent and immutable ledger managed by a network of nodes. However, decentralized systems face challenges in maintaining compliance with evolving legal requirements, which requires immediate and dynamic updates~\cite{toorani2021decentralized}.

\subsubsection*{Legal Constraints in Centralized Systems}

CPKI systems lack agility and fast processes to adapt to diverse and evolving regulatory requirements across jurisdictions. Centralized hierarchical structures can hinder the ability to implement necessary changes quickly in response to new regulations, potentially leading to compliance issues and legal vulnerabilities. Legal liabilities are concentrated within a few CAs, placing a significant burden on these entities and exposing them to substantial risks in the event of failures or breaches~\cite{serrano2019complete}. In decentralized systems, while the distribution of trust can improve resilience, it complicates decision-making processes. Adapting a smart contract to new regulations requires consensus among multiple entities, which may be slower than the response of a single decision point.

\subsubsection*{Legal Responsiveness with Decentralized Systems}

DPKI systems improve legal responsiveness by supporting cross-border legal diversity through alignment with various regulatory frameworks. Reliable and transparent registries enable DPKI systems to navigate legal environments, facilitating this alignment. The distribution of legal responsibilities among multiple entities fosters a collaborative environment, reducing the burden on individual entities to manage digital trust~\cite{Sedlmeir2021DigitalIA}. However, it is important to note that blockchain technology, integral to many DPKI systems, faces challenges in complying with regulations such as the General Data Protection Regulation (GDPR). The inherent immutability and decentralized nature of blockchain can conflict with GDPR requirements, including the right to be forgotten, complicating the legal compliance of distributed ledger-based systems~\cite{tatar2020law}. Additionally, different smart contracts on different nodes introduce potential inconsistencies that need clarification. Cross-border misalignment remains a problem that needs resolution, as it is unclear which legislation will be applied.

The GDPR presents specific challenges to blockchain-based systems due to its requirements for data protection and the right to be forgotten. Decentralized frameworks must ensure that personal data is handled in compliance with GDPR, which may involve designing mechanisms for data minimization and user consent. The European Union's eIDAS~\cite{eidasAmendment} regulation provides a legal framework for electronic identities and trust services. Integrating eIDAS-compliant identity services within a decentralized framework promises to enhance legal recognition and trust across borders. Our proposed framework addresses legal complexities and enhances compliance by incorporating these regulatory considerations.

\subsection{Credential Status Management}

Incorporating dynamic and decentralized credential status management is essential for DPKI. CPKI systems typically manage the credential status through CRLs and the OCSP. Although CRLs and the OCSP are effective, they encounter challenges such as update delays and scalability issues~\cite{korzhitskii2021revocation}. DPKI improves responsibility distribution, but also faces update delays inherent in distributed systems~\cite{shi2022blockchain}.

Decentralized systems support credential statuses like expiration dates, one-time usage, and dynamic status. Dynamic status credentials can be revoked and unrevoked at any time, with or without providing a reason. However, these systems face challenges in ensuring the privacy of credential issuers, holders, and verifiers. Preventing profiling and securing the privacy of all stakeholders is crucial~\cite{9566179}. Accumulator-based and zero-knowledge-based approaches, while promising, still face challenges in maintaining the privacy of all parties involved and require further optimization for widespread use in decentralized credential status management~\cite{helminger2021multi}. Specifically, accumulators must address efficiency issues related to witness generation and updates. Zero-knowledge proofs must be optimized to reduce computational overhead and ensure scalability in large-scale decentralized systems.

Based on our comparative analysis of CPKI and DPKI, we derive the following requirements for an effective Decentralized Credential Status Management system. Ranked from highest priority (\textit{R1}) to lowest priority (\textit{R10}), these requirements address the aspects of a decentralized credential system, focusing on security, privacy, scalability, and user adoption.

\begin{enumerate}
    \item[\textit{R1}] Privacy of credential issuers, holders, and verifiers using privacy-preserving technologies.
    
    \item[\textit{R2}] Scalability to handle growing transactions without performance degradation.

    \item[\textit{R3}] Interoperability with standardized credential issuance, presentation, and verification protocols.
        
    \item[\textit{R4}] Security through cryptographically solid techniques for data protection during storage, transmission, and computation.
    
    \item[\textit{R5}] Decentralized storage solutions for secure credential status information and resilient data retrieval systems.
    
    \item[\textit{R6}] Efficiency in real-time updates to credential status to address latency issues and optimize cryptographic operations.
    
    \item[\textit{R7}] Secure and efficient communication protocols for data transmission and real-time updates.
    
    \item[\textit{R8}] Simplified lifecycle management of credentials, including issuance, renewal, and revocation, with support for dynamic status credentials.
    
    \item[\textit{R9}] Governance frameworks to ensure compliance with diverse and evolving legal requirements across jurisdictions.
    
    \item[\textit{R10}] User-friendly interfaces and platforms for managing decentralized identities and credentials to facilitate adoption.
\end{enumerate}

In summary, the transition from CPKI to DPKI provides a robust and efficient approach to managing digital trust and verifiable credentials by mitigating the vulnerabilities and inefficiencies inherent in centralized systems. DPKI offers a promising digital credential and trust management solution in an increasingly interconnected world, leveraging decentralized technologies that enhance resistance to attack and immutability. However, DPKI continues to face challenges related to efficiency and performance. This transition sets the stage for further exploration of dynamic and decentralized credential status management, as discussed in the following section.

\section{Framework for Dynamic and Decentralized Credential Status Management}

Decentralized systems need innovative credential management approaches. Unlike traditional CPKI systems, DPKI leverages distributed ledgers, consensus algorithms, and smart contracts to improve security and transparency, although improvements in scalability and efficiency are still required.

\begin{figure}[!h]
    \centering
    \includegraphics[width=\linewidth]{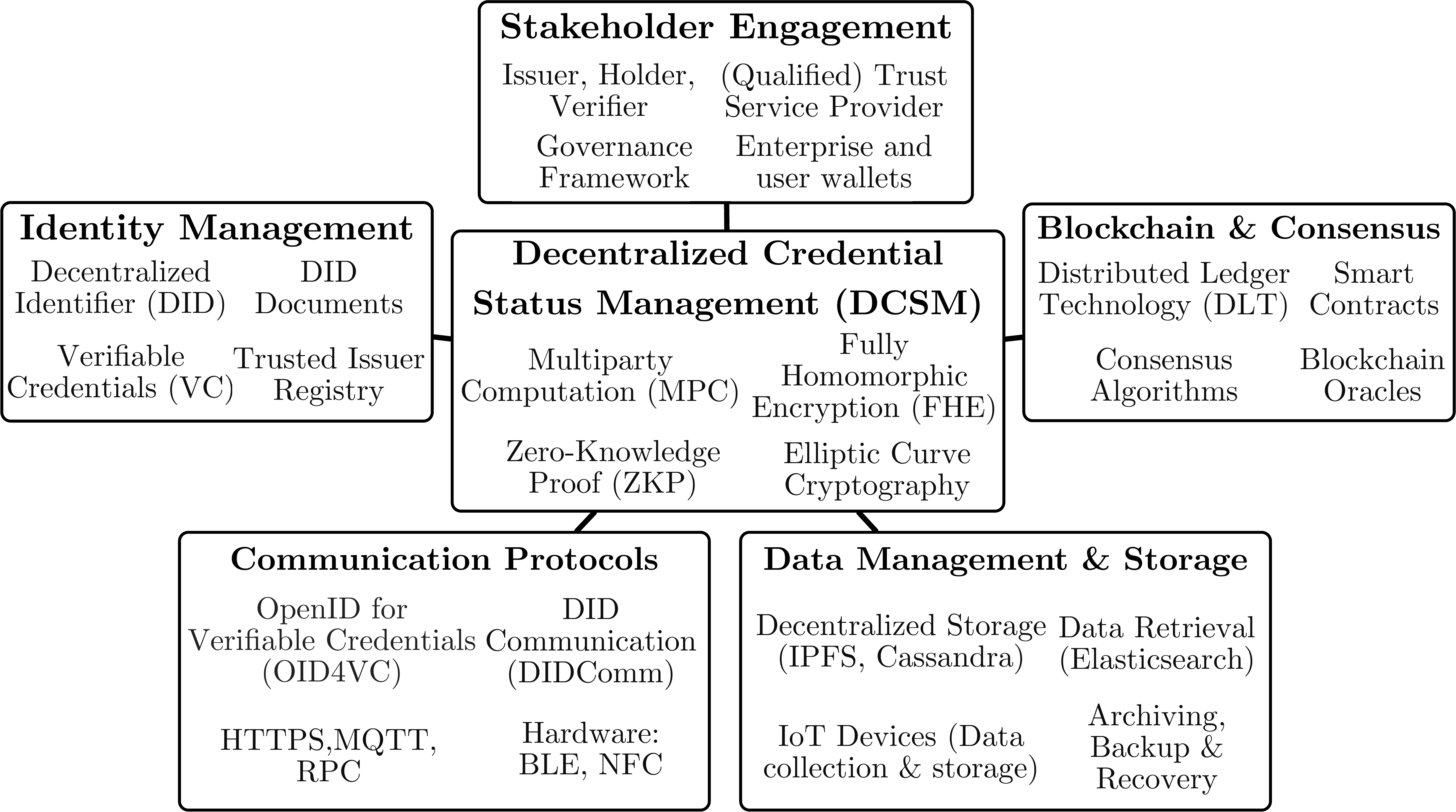}
    \caption{Key components of a DCSM framework include identity management, stakeholder engagement, blockchain and consensus mechanisms, communication and data transport, and data management and storage. These components collaboratively ensure secure, scalable, and efficient management of digitally verifiable credentials in a decentralized framework.}
    \label{fig:framework-decentralized}
\end{figure}

We propose a user-centered and privacy-preserving DPKI framework using decentralized technologies aligned with the SSI paradigm~\cite{10246272}. Figure~\ref{fig:framework-decentralized} highlights DCSM, focusing on managing verifiable credential status. The following sections detail the framework's components and interactions, aligning with the requirements for effective DCSM. In addition, we list the requirements that are addressed by each component \textit{(R1-R10)}.

\subsection{Identity Management}
Decentralized identity management enables user-centric digital identities within the SSI framework, allowing control without central authorities. SSI’s components, DIDs, and VCs, provide secure identification and tamper-proof attestations such as identity documents and qualifications. VCs support privacy-preserving verifications, ensuring minimal disclosure~\textit{(R1, R3, R4, R5)}~\cite{brunner2020did}. However, a root authority for the actual identity, beyond the identifiers, is essential to ensure trust in centralized and decentralized systems. Although decentralizing attributes and the verification process is possible, lacking a trusted root authority for the core identity information would leave the verification chain unsupported.

However, there is a need to address substantial gaps and challenges in decentralized identity management, particularly in the management of credential status. Current credential status management solutions, such as blockchain-based accumulator solutions and CRLs for credential revocation, involve privacy and scalability trade-offs~\cite{baldimtsi2017accumulators}. A primary challenge is maintaining the privacy of issuers, holders, and verifiers during the verification and issuance processes while ensuring transparency and security. This prevents profiling and tracking of the entities involved.

\subsection{Stakeholder Engagement} 

Effective governance and stakeholder participation are critical to the success of DCSM. Governance frameworks, such as eIDAS 2.0 and the European Blockchain Services Infrastructure (EBSI), establish regulatory and operational standards that ensure interoperability, security, and trust between stakeholders \textit{(R4, R5, R7, R9)}~\cite{giannopoulou2023digital}.

Key stakeholders within decentralized ecosystems include issuers, holders, verifiers, and qualified trust service providers. User wallets store authentic data, and enterprise wallets issue signed VCs, managing stakeholder interactions. Wallets ensure privacy and security and support the operational integrity of the decentralized network. Verifier services validate credentials and maintain the trustworthiness of the ecosystem. Addressing stakeholder privacy and ensuring compliance with regulatory environments is essential to establish a robust, privacy-preserving, and user-centric DCSM framework~\cite{lips2022re}.

The challenges of stakeholder participation include ensuring the anonymity and privacy of all parties involved in diverse and evolving regulatory environments. Implementing DAOs can improve the resilience and transparency of the DPKI system by enabling collaborative decision-making and governance~\cite{santana2022blockchain}. By addressing these challenges through robust governance frameworks and stakeholder engagement strategies, DCSM can provide a secure, privacy-preserving, and user-centric framework, thus enabling the full potential of DPKI.

\subsection{Blockchain and Consensus}

Blockchain technology enables DPKI by storing DIDs and associated DID Documents in an immutable ledger. The blockchain records DIDs linked to DID Documents containing public keys and service endpoints. It also maintains revocation registries and logs DID operations, such as updates and status changes. This setup verifies the integrity and authenticity of verifiable credentials without storing the transaction details of the issuance. Blockchains' ledger nodes are synchronized through consensus protocols, ensuring consistent and reliable updates. The decentralized nature of blockchains ensures that no single entity can control or manipulate the system. Blockchain immutability strengthens security against tampering but does not prevent false information if an attacker compromises an authorized writer's credentials. In such cases, the false data would be written in the ledger and replicated across all copies, highlighting a vulnerability in credential security. Furthermore, node synchronization can introduce latency, and the need for atomic propagation, where no information is visible until the entire process is complete, can affect efficiency~\textit{(R1, R5, R6, R8)}~\cite{sporny2022decentralized}.

Consensus algorithms, such as Proof of Work (PoW), Proof of Stake (PoS), and Proof of Authority (PoA), validate and secure transactions. These transactions include registering DIDs, updating DID Documents, and revocation registries. Consensus algorithms ensure that network participants verify and reach a consensus on all transactions before writing them to the ledger. PoW relies on computational power to solve cryptographic puzzles, while PoS selects validators based on the number of tokens they hold and are willing to stake as collateral. PoA selects validators based on their identity and reputation within the network. These consensus mechanisms maintain the integrity of the ledger by preventing fraudulent activities and ensuring the security and reliability of the blockchain~\cite{bach2018comparative}.

Smart contracts are digital contracts encoded with predefined rules that automate blockchain transactions, such as credential revocation and verification. Without intermediaries, smart contracts facilitate tamper-proof blockchain transactions and credential operations. The transparency of blockchain technology ensures that all transactions are visible and verifiable by network participants, increasing trust and accountability~\cite{buterin2014next}. Integrating smart contracts with DPKI allows the management of digital credentials and trust in a decentralized manner. In addition, oracles, which are services that provide external data feeds into smart contracts, improve the reliability of data sources~\cite{pasdar2023connect}.

\subsection{Data Management and Storage}
Decentralized storage solutions, such as IPFS~\cite{nyaletey2019blockipfs} and Cassandra~\cite{cassandra2014apache}, are essential to securely store credential status information in DCSM. Retrieval systems such as Elasticsearch~\cite{arer2022efficient} enable efficient data access. These systems must be resilient to attacks and support high throughput. Separating data management from blockchain storage ensures operational efficiency. IPFS provides immutable and decentralized storage, while Cassandra offers fault-tolerant and scalable storage. Cassandra contrasts with IPFS in terms of decentralization. IPFS is a decentralized storage solution that does not require mutual trust among nodes, while Cassandra is a distributed database that relies on trust among its nodes. In general, key challenges for data management and storage include the deployment of consortium-controlled data clusters and the effective management of data access control~\textit{(R1, R5, R6, R9)}~\cite{liu2021consortium}. 

\subsection{Communication Protocols}
Communication protocols are necessary for the reliable and secure exchange of credential information within a decentralized network. In the context of DCSM, protocols such as OpenID for Verifiable Credentials (OID4VC) and Decentralized Identifier Communication (DIDComm) provide methods for securely issuing, presenting, and verifying credentials~\textit{(R4, R10)}~\cite{Lux2020DistributedLedgerbasedAW}. These protocols enable secure communication between issuers, holders, and verifiers, ensuring the authenticity and confidentiality of credential exchanges and seamless stakeholder interactions.

Application layer protocols such as HTTPS and MQTT facilitate data transmission and contribute to data integrity. HTTPS, which is HTTP over TLS connections, and MQTT, which can operate over TLS, rely on underlying transport layer protocols such as SSL / TLS and QUIC for encryption and security~\cite{kumar2019implementation}. Remote Procedure Calls (RPC) facilitate communication between system components by allowing one program to request a service from a program located on another computer in the network. This integration helps coordinate credential-related operations~\cite{birrell1984implementing}. Hardware solutions like Bluetooth Low Energy (BLE) and Near Field Communication (NFC) support secure communication for IoT devices, enhancing DCSM versatility and reach. These protocols support high throughput for dynamic credential management, enabling real-time updates and interactions~\textit{(R6, R7)}~\cite{cerruela2016state}.

\subsection{Decentralized Credential Status Management (DCSM)}

Until now, the discussion has focused on established components in identity solutions, such as the Hyperledger and EBSI ecosystems. However, additional advancements are necessary to achieve the full potential of DPKI and ensure the privacy of all stakeholders within decentralized identity management. Maintaining stakeholder privacy prevents tracking and profiling, common in the advertising industry~\cite{raschke2019towards}. Decentralized systems must support various credential statuses, including expiration dates, one-time usage, and dynamic statuses that can be revoked or unrevoked anytime.

Although accumulator-based and ZKP-based approaches offer promising solutions for privacy protection, they require optimization to be practical for widespread use in DCSM. Accumulators and ZKPs ensure that stakeholders can verify credentials without revealing sensitive information, thus protecting privacy and preventing profiling. However, their downsides include high computational complexity and significant resource consumption, leading to inefficiencies and scalability challenges~\textit{(R1, R2, R3, R8)}~\cite{yang2020zero}. The following technologies hold promise for advancing DCSM and realizing the full potential of DPKI.

\subsubsection*{Multiparty Computation (MPC)} is a cryptographic technique that enables multiple parties to collaboratively compute a function over their private inputs without revealing the underlying data. In the context of DCSM, integrating MPC facilitates secure and distributed credential issuance and (un)revocation processes, ensuring that no single entity has complete control over the credential lifecycle. MPC is particularly relevant in DCSM, where trust is decentralized and no single entity has complete visibility and control over the processed data. Furthermoe, MPC enhances trust in DCSM by distributing the processing of credential data among multiple parties, thus preventing any single point of control or failure. However, the practical deployment of MPC in DCSM faces computational and communication challenges due to its complexity. Ongoing research focuses on improving the efficiency and scalability of MPC protocols to enable seamless integration with high-throughput blockchain environments, bridging the gap between theoretical guarantees and practical applications~\textit{(R1, R2, R6, R7)}~\cite{zhao2019secure}. 

\subsubsection*{Fully Homomorphic Encryption (FHE)} is an encryption scheme that allows computations on encrypted data without decryption. FHE is crucial in preserving the privacy of sensitive credential data throughout its lifecycle, from issuance and storage to verification and (un)revocation. Privacy preservation with FHE ensures that credential data remains secure even during processing, thereby enhancing the overall security framework of DCSM. By leveraging FHE, DCSM implementations can ensure that credential status data remains encrypted while allowing necessary computations and updates without requiring any party to access the credentials' status raw data. Combining FHE with MPC, DCSM implementations can enable secure and private computation of encrypted credential data across multiple parties, ensuring that no single entity can access the raw data. Thus, the combination of MPC and FHE provides a privacy-preserving toolset for decentralized systems, enabling distributed credential issuance, verification, and (un)revocation. However, the practical implementation of FHE in DCSM implementations faces challenges primarily related to computational complexity. Ongoing research efforts aim to optimize FHE algorithms, develop more efficient implementations, and explore partially homomorphic encryption (PHE) schemes to mitigate these challenges~\textit{(R1, R2, R4, R6)}~\cite{acar2018survey}.

\subsubsection*{Zero-knowledge proofs (ZKP)} are cryptographic primitives that allow the verification of claims without revealing additional information beyond the validity of the claim. In DCSM, ZKP plays a cruical role in enabling privacy-preserving credential verification, allowing users to prove possession of a credential without disclosing sensitive information. ZKP enhances user trust by ensuring that only the necessary proof is shared, keeping the underlying data private. Integrating ZKP with blockchain technology ensures the security and immutability of the verification process, as proofs are recorded on a tamper-proof ledger. However, the practical implementation of ZKP in DCSM faces proof and computational efficiency challenges. Ongoing research efforts aim to optimize proof construction and verification times, reduce the on-chain computation load of proofs, and improve the user-friendliness of ZKP-based systems. Exploring novel cryptographic primitives, such as zk-SNARKs, zk-STARKs, and ZoKrates, which remove the need for a trusted setup, holds promise for the future of ZKPs in DCSM~\textit{(R1, R3, R4, R5)}~\cite{DBLP:conf/ithings/EberhardtT18}.

\subsubsection*{Elliptic Curve Cryptography (ECC)} provides robust security with shorter key lengths than traditional cryptographic systems, such as RSA. Shorter key lengths enhance the processing speed and reduce computational overhead, making ECC suitable for DCSM applications, especially on devices with limited resources. ECC enables efficient cryptographic operations, which is critical for maintaining the performance of DCSM across diverse devices and environments. Using ECC, DCSM can improve the cryptographic security of credential transactions while ensuring efficient operational performance across the decentralized network~\textit{(R1, R2, R4, R7)}~\cite{hankerson2021elliptic}.

\subsection{Practical Implementation Examples}

Several real-world scenarios can illustrate the practical implementation of DCSM. A systematic review by Agbo et al.~\cite{agbo2019blockchain} presents challenges and opportunities for decentralized healthcare applications to enhance the security and privacy of patient records. Patient records could be verifiable credentials, and our proposed DCSM framework could ensure secure management. By leveraging blockchain technology and advanced cryptographic techniques, healthcare providers can securely manage patient data and ensure access is restricted to authorized entities, complying with stringent privacy regulations like the Health Insurance Portability and Accountability Act (HIPAA)~\cite{mbonihankuye2019healthcare}.

In the education sector, DCSM streamlines the verification of academic qualifications~\cite{bhaskar2021blockchain},~\cite{herbke2022elmo2eds}. Universities can issue and verify verifiable credentials securely, enabling education institutions and employers to authenticate academic records without intermediaries. This process enhances verification efficiency and ensures the integrity and authenticity of educational records, fostering trust among stakeholders .

These examples demonstrate the proposed framework's practical applicability and effectiveness in enhancing security, privacy, and trust across various domains.

\section{Discussion}
Integrating SSI components, such as DIDs and VCs, with ZKP, MPC, and FHE forms a robust DCSM framework. The presented framework allows secure management, verification, and encryption of credential status, with applications across sectors including education, healthcare, finance, and government services. Blockchain technology provides benefits such as immutability, decentralized trust, and transparency, which are essential to improve digital trust and credential management.

Our framework employs several strategies and techniques to address scalability and performance challenges. Sharding divides on-chain and off-chain storage into smaller, more manageable segments, each capable of processing transactions independently. Sharding could improve the overall throughput and reduce latency~\cite{dang2019towards}. Layer-2 solutions, such as state channels and sidechains, promise to improve the throughput of off-chain transactions, reducing the load on the main blockchain and increasing transaction speeds~\cite{gangwal2023survey}.

However, DCSM deployment is limited by the computational complexity of ZKP, MPC, and FHE, restricting scalability and efficiency. Furthermore, integrating these cryptographic techniques with SSI requires the development of standardized protocols and interfaces to ensure interoperability. The Decentralized Identity Foundation~\cite{dif2024didregistration}, the World Wide Web Consortium~\cite{w3cccg2024didresolution}, and the European Union~\cite{roth2023blockchain} are conducting leading research in this domain.

\section{Future Work}
Future work will focus on optimizing the computational efficiency of ZKP, MPC, and FHE to improve scalability. Emerging technologies such as sharding~\cite{DBLP:journals/access/YuWYNZL20} and Layer-2~\cite{DBLP:conf/isda/PandeyFBT21} solutions offer promising avenues to overcome scalability limitations in blockchain-based DCSM implementations. Developing standardized protocols and interfaces is crucial for integrating these technologies with SSI frameworks. In addition, practical deployment studies are needed to explore real-world applications and address regulatory challenges associated with decentralized systems. Focusing on these areas can improve the robustness and practicality of DCSM systems, meeting the evolving needs of digital credential management.

\section*{Acknowledgment}
This work was supported by the European Union’s Digital Europe Program (DIGITAL) research and innovation program under grant agreement number 101102743 (TRACE4EU). The authors thank the Trace4EU project partners for their invaluable contributions and the cheqd team for their support and expertise.

\bibliographystyle{IEEEtran}
\bibliography{references.bib}

\begin{thebibliography}{10}
\providecommand{\url}[1]{#1}
\csname url@samestyle\endcsname
\providecommand{\newblock}{\relax}
\providecommand{\bibinfo}[2]{#2}
\providecommand{\BIBentrySTDinterwordspacing}{\spaceskip=0pt\relax}
\providecommand{\BIBentryALTinterwordstretchfactor}{4}
\providecommand{\BIBentryALTinterwordspacing}{\spaceskip=\fontdimen2\font plus
\BIBentryALTinterwordstretchfactor\fontdimen3\font minus \fontdimen4\font\relax}
\providecommand{\BIBforeignlanguage}[2]{{%
\expandafter\ifx\csname l@#1\endcsname\relax
\typeout{** WARNING: IEEEtran.bst: No hyphenation pattern has been}%
\typeout{** loaded for the language `#1'. Using the pattern for}%
\typeout{** the default language instead.}%
\else
\language=\csname l@#1\endcsname
\fi
#2}}
\providecommand{\BIBdecl}{\relax}
\BIBdecl

\bibitem{prins2011diginotar}
J.~R. Prins and B.~U. Cybercrime, ``{Diginotar certificate authority breach “operation black tulip”},'' \emph{Fox-IT, November}, vol.~18, 2011.

\bibitem{DBLP:conf/pam/GasserHHKHC18}
O.~Gasser, B.~Hof, M.~Helm, M.~Korczynski, R.~Holz, and G.~Carle, ``{In Log We Trust: Revealing Poor Security Practices with Certificate Transparency Logs and Internet Measurements},'' in \emph{{PAM}}, ser. Lecture Notes in Computer Science, vol. 10771.\hskip 1em plus 0.5em minus 0.4em\relax Springer, 2018, pp. 173--185.

\bibitem{daswani2021equifax}
N.~Daswani, M.~Elbayadi, N.~Daswani, and M.~Elbayadi, ``The equifax breach,'' \emph{Big Breaches: Cybersecurity Lessons for Everyone}, pp. 75--95, 2021.

\bibitem{9663537}
W.~Li, W.~Meng, and L.~F. Kwok, ``Surveying trust-based collaborative intrusion detection: State-of-the-art, challenges and future directions,'' \emph{IEEE Communications Surveys \& Tutorials}, vol.~24, no.~1, pp. 280--305, 2022.

\bibitem{DBLP:conf/sp/MatsumotoR17}
S.~Matsumoto and R.~M. Reischuk, ``{{IKP:} Turning a {PKI} Around with Decentralized Automated Incentives},'' in \emph{{IEEE} Symposium on Security and Privacy}.\hskip 1em plus 0.5em minus 0.4em\relax {IEEE} Computer Society, 2017, pp. 410--426.

\bibitem{preukschat2021self}
A.~Preukschat and D.~Reed, \emph{{Self-sovereign identity}}.\hskip 1em plus 0.5em minus 0.4em\relax Manning Publications, 2021.

\bibitem{1055638}
W.~Diffie and M.~Hellman, ``{New directions in cryptography},'' \emph{IEEE Transactions on Information Theory}, vol.~22, no.~6, pp. 644--654, 1976.

\bibitem{10.1145/359340.359342}
R.~L. Rivest, A.~Shamir, and L.~M. Adleman, ``{A Method for Obtaining Digital Signatures and Public-Key Cryptosystems (Reprint)},'' \emph{Commun. {ACM}}, vol.~26, no.~1, pp. 96--99, 1983.

\bibitem{merkle22method}
R.~C. Merkle, ``{Method of providing digital signatures},'' \emph{URL: https://www.google.com/patents/US4309569 (cit. on p. 22)}, 1979.

\bibitem{DBLP:conf/stoc/GoldwasserMR85}
S.~Goldwasser, S.~Micali, and C.~Rackoff, ``{The Knowledge Complexity of Interactive Proof-Systems (Extended Abstract)},'' in \emph{{STOC}}.\hskip 1em plus 0.5em minus 0.4em\relax {ACM}, 1985, pp. 291--304.

\bibitem{x509}
C.C.I.T.T., ``{Recommendation X .509, The Directory - Authentication Framework},'' \emph{C.C.I.T.T}, December 1988.

\bibitem{chaum1983blind}
D.~Chaum, ``{Blind signatures for untraceable payments},'' in \emph{Advances in Cryptology: Proceedings of Crypto 82}.\hskip 1em plus 0.5em minus 0.4em\relax Springer, 1983, pp. 199--203.

\bibitem{dierks1999tls}
T.~Dierks and C.~Allen, ``{The TLS protocol version 1.0},'' Internet Engineering Task Force (IETF), Tech. Rep., 1999.

\bibitem{DBLP:conf/fc/McDanielR00}
P.~D. McDaniel and A.~D. Rubin, ``{A Response to ''Can We Eliminate Certificate Revocation Lists?''},'' in \emph{Financial Cryptography}, ser. Lecture Notes in Computer Science, vol. 1962.\hskip 1em plus 0.5em minus 0.4em\relax Springer, 2000, pp. 245--258.

\bibitem{DBLP:conf/mm/Wohlmacher00}
P.~Wohlmacher, ``{Digital certificates: a survey of revocation methods},'' in \emph{{ACM} Multimedia Workshops}.\hskip 1em plus 0.5em minus 0.4em\relax {ACM} Press, 2000, pp. 111--114.

\bibitem{DBLP:books/daglib/0084694}
S.~L. Garfinkel, \emph{{{PGP} - pretty good privacy: encryption for everyone {(2.} ed.)}}.\hskip 1em plus 0.5em minus 0.4em\relax O'Reilly, 1995.

\bibitem{Barnes2019AutomaticCM}
R.~L. Barnes, J.~Hoffman-Andrews, D.~McCarney, and J.~Kasten, ``{Automatic Certificate Management Environment (ACME)},'' \emph{RFC}, vol. 8555, pp. 1--95, 2019.

\bibitem{10.1145/2659897}
\BIBentryALTinterwordspacing
B.~Laurie, ``{Certificate transparency},'' \emph{Commun. ACM}, vol.~57, no.~10, p. 40–46, sep 2014. [Online]. Available: \url{https://doi.org/10.1145/2659897}
\BIBentrySTDinterwordspacing

\bibitem{10.1145/2815675.2815685}
Y.~Liu, W.~Tome, L.~Zhang, D.~R. Choffnes, D.~Levin, B.~M. Maggs, A.~Mislove, A.~Schulman, and C.~Wilson, ``{An End-to-End Measurement of Certificate Revocation in the Web's {PKI}},'' in \emph{Internet Measurement Conference}.\hskip 1em plus 0.5em minus 0.4em\relax {ACM}, 2015, pp. 183--196.

\bibitem{finney2004rpow}
H.~Finney, ``{Rpow-reusable proofs of work},'' \emph{Internet: https://cryptome.org/rpow.htm}, 2004.

\bibitem{nakamoto2008bitcoin}
S.~Nakamoto, ``{Bitcoin: A peer-to-peer electronic cash system},'' \emph{Decentralized business review}, 2008.

\bibitem{DBLP:journals/fgcs/ZhengXDCCWI20}
Z.~Zheng, S.~Xie, H.~Dai, W.~Chen, X.~Chen, J.~Weng, and M.~Imran, ``{An overview on smart contracts: Challenges, advances and platforms},'' \emph{Future Gener. Comput. Syst.}, vol. 105, pp. 475--491, 2020.

\bibitem{DBLP:journals/tcss/WangDLYOW19}
S.~Wang, W.~Ding, J.~Li, Y.~Yuan, L.~Ouyang, and F.~Wang, ``{Decentralized Autonomous Organizations: Concept, Model, and Applications},'' \emph{{IEEE} Trans. Comput. Soc. Syst.}, vol.~6, no.~5, pp. 870--878, 2019.

\bibitem{tobin2016inevitable}
A.~Tobin and D.~Reed, ``{The inevitable rise of self-sovereign identity},'' \emph{The Sovrin Foundation}, vol.~29, no. 2016, p.~18, 2016.

\bibitem{reed2020decentralized}
M.~Sporny, D.~Longley, M.~Sabadello, D.~Reed, O.~Steele, and C.~Allen, ``{Decentralized identifiers (dids) v1. 0},'' \emph{W3C Recommendation}, 2022.

\bibitem{DBLP:journals/corr/abs-2006-04754}
Z.~A. Lux, D.~Thatmann, S.~Zickau, and F.~Beierle, ``{Distributed-Ledger-based Authentication with Decentralized Identifiers and Verifiable Credentials},'' \emph{CoRR}, vol. abs/2006.04754, 2020.

\bibitem{Fromknecht2014ADP}
C.~Fromknecht, D.~Velicanu, and S.~Yakoubov, ``{A Decentralized Public Key Infrastructure with Identity Retention},'' \emph{IACR Cryptol. ePrint Arch.}, vol. 2014, p. 803, 2014.

\bibitem{axon2016pb}
L.~Axon and M.~Goldsmith, ``{{PB-PKI:} {A} Privacy-aware Blockchain-based {PKI}},'' in \emph{{SECRYPT}}.\hskip 1em plus 0.5em minus 0.4em\relax SciTePress, 2017, pp. 311--318.

\bibitem{vanin2022blockchain}
F.~N. d.~S. Vanin, L.~M. Policarpo, R.~d.~R. Righi, S.~M. Heck, V.~F. da~Silva, J.~Goldim, and C.~A. da~Costa, ``A blockchain-based end-to-end data protection model for personal health records sharing: a fully homomorphic encryption approach,'' \emph{Sensors}, vol.~23, no.~1, p.~14, 2022.

\bibitem{kubilay2019certledger}
M.~Y. Kubilay, M.~S. Kiraz, and H.~A. Mantar, ``{CertLedger: A new PKI model with certificate transparency based on blockchain},'' \emph{Computers \& Security}, vol.~85, pp. 333--352, 2019.

\bibitem{9201400}
G.~Fragkos, C.~Minwalla, J.~Plusquellic, and E.~E. Tsiropoulou, ``Artificially intelligent electronic money,'' \emph{IEEE Consumer Electronics Magazine}, vol.~10, no.~4, pp. 81--89, 2021.

\bibitem{muhle2018survey}
A.~M{\"u}hle, A.~Gr{\"u}ner, T.~Gayvoronskaya, and C.~Meinel, ``{A survey on essential components of a self-sovereign identity},'' \emph{Computer Science Review}, vol.~30, pp. 80--86, 2018.

\bibitem{217541}
H.~Birge{-}Lee, Y.~Sun, A.~Edmundson, J.~Rexford, and P.~Mittal, ``{Bamboozling Certificate Authorities with {BGP}},'' in \emph{{USENIX} Security Symposium}.\hskip 1em plus 0.5em minus 0.4em\relax {USENIX} Association, 2018, pp. 833--849.

\bibitem{chung2018web}
T.~Chung, J.~Lok, B.~Chandrasekaran, D.~R. Choffnes, D.~Levin, B.~M. Maggs, A.~Mislove, J.~P. Rula, N.~Sullivan, and C.~Wilson, ``{Is the Web Ready for {OCSP} Must-Staple?}'' in \emph{Internet Measurement Conference}.\hskip 1em plus 0.5em minus 0.4em\relax {ACM}, 2018, pp. 105--118.

\bibitem{qin2020cecoin}
B.~Qin, J.~Huang, Q.~Wang, X.~Luo, B.~Liang, and W.~Shi, ``{Cecoin: {A} decentralized {PKI} mitigating MitM attacks},'' \emph{Future Gener. Comput. Syst.}, vol. 107, pp. 805--815, 2020.

\bibitem{zheng2017overview}
Z.~Zheng, S.~Xie, H.~Dai, X.~Chen, and H.~Wang, ``{An Overview of Blockchain Technology: Architecture, Consensus, and Future Trends},'' in \emph{BigData Congress}.\hskip 1em plus 0.5em minus 0.4em\relax {IEEE} Computer Society, 2017, pp. 557--564.

\bibitem{papageorgiou2020dpki}
A.~Papageorgiou, A.~Mygiakis, K.~Loupos, and T.~Krousarlis, ``{DPKI: a blockchain-based decentralized public key infrastructure system},'' in \emph{2020 Global Internet of Things Summit (GIoTS)}.\hskip 1em plus 0.5em minus 0.4em\relax IEEE, 2020, pp. 1--5.

\bibitem{9343191}
A.~Abraham, S.~More, C.~Rabensteiner, and F.~Hörandner, ``{Revocable and Offline-Verifiable Self-Sovereign Identities},'' in \emph{2020 IEEE 19th International Conference on Trust, Security and Privacy in Computing and Communications (TrustCom)}, 2020, pp. 1020--1027.

\bibitem{xu2021slimchain}
C.~Xu, C.~Zhang, J.~Xu, and J.~Pei, ``{Slimchain: Scaling blockchain transactions through off-chain storage and parallel processing},'' \emph{Proceedings of the VLDB Endowment}, vol.~14, no.~11, pp. 2314--2326, 2021.

\bibitem{pop00001}
\BIBentryALTinterwordspacing
A.~Fulmer, ``{Trust in Organizational Contexts},'' \emph{Management}, 2019, query date: 2024-06-08 17:27:15. [Online]. Available: \url{http://dx.doi.org/10.1093/obo/9780199846740-0159}
\BIBentrySTDinterwordspacing

\bibitem{geneiatakis2020blockchain}
D.~Geneiatakis, Y.~Soupionis, G.~Steri, I.~Kounelis, R.~Neisse, and I.~N. Fovino, ``{Blockchain Performance Analysis for Supporting Cross-Border E-Government Services},'' \emph{{IEEE} Trans. Engineering Management}, vol.~67, no.~4, pp. 1310--1322, 2020.

\bibitem{braun2013potential}
J.~Braun and G.~Rynkowski, ``{The potential of an individualized set of trusted cas: Defending against ca failures in the web pki},'' in \emph{2013 International Conference on Social Computing}.\hskip 1em plus 0.5em minus 0.4em\relax IEEE, 2013, pp. 600--605.

\bibitem{toorani2021decentralized}
M.~Toorani and C.~Gehrmann, ``{A decentralized dynamic pki based on blockchain},'' in \emph{Proceedings of the 36th annual ACM symposium on applied computing}, 2021, pp. 1646--1655.

\bibitem{serrano2019complete}
N.~Serrano, H.~Hadan, and L.~J. Camp, ``{A complete study of PKI (PKI’s Known Incidents)},'' in \emph{TPRC47: The 47th Research Conference on Communication, Information and Internet Policy}, 2019.

\bibitem{Sedlmeir2021DigitalIA}
J.~Sedlmeir, R.~Smethurst, A.~Rieger, and G.~Fridgen, ``{Digital Identities and Verifiable Credentials},'' \emph{Business \& Information Systems Engineering}, vol.~63, pp. 603--613, 2021.

\bibitem{tatar2020law}
U.~Tatar, Y.~Gokce, and B.~Nussbaum, ``{Law versus technology: Blockchain, GDPR, and tough tradeoffs},'' \emph{Computer Law \& Security Review}, vol.~38, p. 105454, 2020.

\bibitem{eidasAmendment}
C.~o. t. E.~U. European~Parliament, ``Eu regulation 32024r1183,'' \url{https://eur-lex.europa.eu/legal-content/EN/TXT/?uri=celex%3A32024R1183}, accessed: 2024-07-20.

\bibitem{korzhitskii2021revocation}
N.~Korzhitskii and N.~Carlsson, ``{Revocation Statuses on the Internet},'' in \emph{{PAM}}, ser. Lecture Notes in Computer Science, vol. 12671.\hskip 1em plus 0.5em minus 0.4em\relax Springer, 2021, pp. 175--191.

\bibitem{shi2022blockchain}
J.~Shi, X.~Zeng, and R.~Han, ``{A blockchain-based decentralized public key infrastructure for information-centric networks},'' \emph{Information}, vol.~13, no.~5, p. 264, 2022.

\bibitem{9566179}
F.~Béres, I.~A. Seres, A.~A. Benczúr, and M.~Quintyne-Collins, ``{Blockchain is Watching You: Profiling and Deanonymizing Ethereum Users},'' in \emph{2021 IEEE International Conference on Decentralized Applications and Infrastructures (DAPPS)}, 2021, pp. 69--78.

\bibitem{helminger2021multi}
L.~Helminger, D.~Kales, S.~Ramacher, and R.~Walch, ``{Multi-party revocation in sovrin: Performance through distributed trust},'' in \emph{Cryptographers’ Track at the RSA Conference}.\hskip 1em plus 0.5em minus 0.4em\relax Springer, 2021, pp. 527--551.

\bibitem{10246272}
H.~Yildiz, A.~Küpper, D.~Thatmann, S.~Göndör, and P.~Herbke, ``{Toward Interoperable Self-Sovereign Identities},'' \emph{IEEE Access}, vol.~11, pp. 114\,080--114\,116, 2023.

\bibitem{brunner2020did}
C.~Brunner, U.~Gallersd{\"{o}}rfer, F.~Knirsch, D.~Engel, and F.~Matthes, ``{{DID} and {VC:} Untangling Decentralized Identifiers and Verifiable Credentials for the Web of Trust},'' in \emph{{ICBTA}}.\hskip 1em plus 0.5em minus 0.4em\relax {ACM}, 2020, pp. 61--66.

\bibitem{baldimtsi2017accumulators}
F.~Baldimtsi, J.~Camenisch, M.~Dubovitskaya, A.~Lysyanskaya, L.~Reyzin, K.~Samelin, and S.~Yakoubov, ``{Accumulators with Applications to Anonymity-Preserving Revocation},'' 2017, p.~43.

\bibitem{giannopoulou2023digital}
A.~Giannopoulou, ``{Digital identity infrastructures: a critical approach of self-sovereign identity},'' \emph{Digital Society}, vol.~2, no.~2, p.~18, 2023.

\bibitem{lips2022re}
S.~Lips, N.~Vinogradova, R.~Krimmer, and D.~Draheim, ``{Re-Shaping the {EU} Digital Identity Framework},'' in \emph{{DG.O}}.\hskip 1em plus 0.5em minus 0.4em\relax {ACM}, 2022, pp. 13--21.

\bibitem{santana2022blockchain}
C.~Santana and L.~Albareda, ``{Blockchain and the emergence of Decentralized Autonomous Organizations (DAOs): An integrative model and research agenda},'' \emph{Technological Forecasting and Social Change}, vol. 182, p. 121806, 2022.

\bibitem{sporny2022decentralized}
M.~Sporny, D.~Longley, M.~Sabadello, D.~Reed, O.~Steele, and C.~Allen, ``{Decentralized Identifiers (DIDs) v1. 0 Core architecture, data model, and representations},'' \emph{W3C Working Draft}, 2022.

\bibitem{bach2018comparative}
L.~M. Bach, B.~Mihaljevic, and M.~Zagar, ``Comparative analysis of blockchain consensus algorithms,'' in \emph{{MIPRO}}.\hskip 1em plus 0.5em minus 0.4em\relax {IEEE}, 2018, pp. 1545--1550.

\bibitem{buterin2014next}
V.~Buterin \emph{et~al.}, ``{A next-generation smart contract and decentralized application platform},'' \emph{white paper}, vol.~3, no.~37, pp. 2--1, 2014.

\bibitem{pasdar2023connect}
A.~Pasdar, Y.~C. Lee, and Z.~Dong, ``{Connect api with blockchain: A survey on blockchain oracle implementation},'' \emph{ACM Computing Surveys}, vol.~55, no.~10, pp. 1--39, 2023.

\bibitem{nyaletey2019blockipfs}
E.~Nyaletey, R.~M. Parizi, Q.~Zhang, and K.~R. Choo, ``Blockipfs - blockchain-enabled interplanetary file system for forensic and trusted data traceability,'' in \emph{Blockchain}.\hskip 1em plus 0.5em minus 0.4em\relax {IEEE}, 2019, pp. 18--25.

\bibitem{cassandra2014apache}
A.~Cassandra, ``{Apache cassandra},'' \emph{Website. Available online at http://planetcassandra. org/what-is-apache-cassandra}, vol.~13, 2014.

\bibitem{arer2022efficient}
M.~M. Arer, P.~M. Dhulavvagol, and S.~Totad, ``{Efficient big data storage and retrieval in distributed architecture using blockchain and ipfs},'' in \emph{2022 IEEE 7th International conference for Convergence in Technology (I2CT)}.\hskip 1em plus 0.5em minus 0.4em\relax IEEE, 2022, pp. 1--6.

\bibitem{liu2021consortium}
S.~Liu and H.~Tang, ``A consortium medical blockchain data storage and sharing model based on {IPFS},'' in \emph{{ICCMB}}.\hskip 1em plus 0.5em minus 0.4em\relax {ACM}, 2021, pp. 147--153.

\bibitem{Lux2020DistributedLedgerbasedAW}
Z.~A. Lux, D.~Thatmann, S.~Zickau, and F.~Beierle, ``{Distributed-Ledger-based Authentication with Decentralized Identifiers and Verifiable Credentials},'' pp. 71--78, 2020.

\bibitem{kumar2019implementation}
P.~Kumar and B.~Dezfouli, ``{Implementation and analysis of QUIC for MQTT},'' \emph{Computer Networks}, vol. 150, pp. 28--45, 2019.

\bibitem{birrell1984implementing}
A.~D. Birrell and B.~J. Nelson, ``{Implementing remote procedure calls},'' \emph{ACM Transactions on Computer Systems (TOCS)}, vol.~2, no.~1, pp. 39--59, 1984.

\bibitem{cerruela2016state}
G.~C. Garc{\'{\i}}a, I.~L. Ruiz, and M.~{\'{A}}. G{\'{o}}mez{-}Nieto, ``{State of the Art, Trends and Future of Bluetooth Low Energy, Near Field Communication and Visible Light Communication in the Development of Smart Cities},'' \emph{Sensors}, vol.~16, no.~11, p. 1968, 2016.

\bibitem{raschke2019towards}
P.~Raschke, S.~Zickau, J.~L. Kr{\"{o}}ger, and A.~K{\"{u}}pper, ``Towards real-time web tracking detection with {T.EX} - the transparency extension,'' in \emph{{APF}}, ser. Lecture Notes in Computer Science, vol. 11498.\hskip 1em plus 0.5em minus 0.4em\relax Springer, 2019, pp. 3--17.

\bibitem{yang2020zero}
X.~Yang and W.~Li, ``{A zero-knowledge-proof-based digital identity management scheme in blockchain},'' \emph{Computers \& Security}, vol.~99, p. 102050, 2020.

\bibitem{zhao2019secure}
C.~Zhao, S.~Zhao, M.~Zhao, Z.~Chen, C.-Z. Gao, H.~Li, and Y.-a. Tan, ``{Secure multi-party computation: theory, practice and applications},'' \emph{Information Sciences}, vol. 476, pp. 357--372, 2019.

\bibitem{acar2018survey}
A.~Acar, H.~Aksu, A.~S. Uluagac, and M.~Conti, ``{A survey on homomorphic encryption schemes: Theory and implementation},'' \emph{ACM Computing Surveys (Csur)}, vol.~51, no.~4, pp. 1--35, 2018.

\bibitem{DBLP:conf/ithings/EberhardtT18}
J.~Eberhardt and S.~Tai, ``{ZoKrates - Scalable Privacy-Preserving Off-Chain Computations},'' in \emph{iThings/GreenCom/CPSCom/SmartData}.\hskip 1em plus 0.5em minus 0.4em\relax {IEEE}, 2018, pp. 1084--1091.

\bibitem{hankerson2021elliptic}
D.~Hankerson and A.~Menezes, ``{Elliptic curve cryptography},'' in \emph{Encyclopedia of Cryptography, Security and Privacy}.\hskip 1em plus 0.5em minus 0.4em\relax Springer, 2021, pp. 1--2.

\bibitem{agbo2019blockchain}
C.~C. Agbo, Q.~H. Mahmoud, and J.~M. Eklund, ``Blockchain technology in healthcare: a systematic review,'' in \emph{Healthcare}, vol.~7, no.~2.\hskip 1em plus 0.5em minus 0.4em\relax MDPI, 2019, p.~56.

\bibitem{mbonihankuye2019healthcare}
S.~Mbonihankuye, A.~Nkunzimana, and A.~Ndagijimana, ``Healthcare data security technology: Hipaa compliance,'' \emph{Wireless communications and mobile computing}, vol. 2019, no.~1, p. 1927495, 2019.

\bibitem{bhaskar2021blockchain}
P.~Bhaskar, C.~K. Tiwari, and A.~Joshi, ``Blockchain in education management: present and future applications,'' \emph{Interactive Technology and Smart Education}, vol.~18, no.~1, pp. 1--17, 2021.

\bibitem{herbke2022elmo2eds}
P.~Herbke and H.~Yildiz, ``Elmo2eds: transforming educational credentials into self-sovereign identity paradigm,'' in \emph{2022 20th International Conference on Information Technology Based Higher Education and Training (ITHET)}.\hskip 1em plus 0.5em minus 0.4em\relax IEEE, 2022, pp. 1--7.

\bibitem{dang2019towards}
H.~Dang, T.~T.~A. Dinh, D.~Loghin, E.-C. Chang, Q.~Lin, and B.~C. Ooi, ``Towards scaling blockchain systems via sharding,'' in \emph{Proceedings of the 2019 international conference on management of data}, 2019, pp. 123--140.

\bibitem{gangwal2023survey}
A.~Gangwal, H.~R. Gangavalli, and A.~Thirupathi, ``A survey of layer-two blockchain protocols,'' \emph{Journal of Network and Computer Applications}, vol. 209, p. 103539, 2023.

\bibitem{dif2024didregistration}
{Markus Sabadello and Cihan Saglam}, ``{DID Registration},'' \url{https://identity.foundation/did-registration/}, 2024, accessed: 2024-06-09.

\bibitem{w3cccg2024didresolution}
{Markus Sabadello and Dmitri Zagidulin}, ``{DID Resolution},'' \url{https://w3c-ccg.github.io/did-resolution/}, 2024, accessed: 2024-06-09.

\bibitem{roth2023blockchain}
T.~Roth, A.~Stohr, J.~Amend, G.~Fridgen, and A.~Rieger, ``{Blockchain as a driving force for federalism: {A} theory of cross-organizational task-technology fit},'' \emph{Int. J. Inf. Manag.}, vol.~68, p. 102476, 2023.

\bibitem{DBLP:journals/access/YuWYNZL20}
G.~Yu, X.~Wang, K.~Yu, W.~Ni, J.~A. Zhang, and R.~P. Liu, ``Survey: Sharding in blockchains,'' \emph{{IEEE} Access}, vol.~8, pp. 14\,155--14\,181, 2020.

\bibitem{DBLP:conf/isda/PandeyFBT21}
A.~A. Pandey, T.~F. Fernandez, R.~Bansal, and A.~K. Tyagi, ``Maintaining scalability in blockchain,'' in \emph{{ISDA}}, ser. Lecture Notes in Networks and Systems, vol. 418.\hskip 1em plus 0.5em minus 0.4em\relax Springer, 2021, pp. 34--45.

\end{thebibliography}

\end{document}